Author: Mladen Georgiev (Institute of Solid State Physics, Bulgarian Academy of Sciences 1784 Sofia, Bulgaria)

Vibronic polarons (charge carriers coupled to their produced Jahn-Teller distortions) have been extended for some time in relevance to the buildup of matter and the breakdown of supersymmetry at the early stages of universe, as well as to a number of well established though not fully understood phenomena of everyday life, such as helical propagation, chirality, colossal magnetoresistance, etc. We review some of the previous advances and suggest new ones. Many of our predictions can be checked by LHC experiments.

## 1. The vibronic polaron

Vibronic polarons (*vp*) have been introduced for the first time by Thomas *et al.* back in the early 1980s in a variational ansatz study [1]. However, they chose a controversial coupling Hamiltonian which reduced their species to Holstein polarons. The problem was tackled later by Brown *et al.* in 1999 hopefully on a more convincing basis [2]. While Thomas *et al.* considered a local deformation of the Jahn-Teller (JT) type, later developments have extended the species by introducing a pseudo-Jahn-Teller (pJT) coupling to obtain a polaron. Accordingly, we now define the *vp* as a charge carrier coupled to its produced pJT distortion. As we shall see shortly, the pJT extension brings about a number of novel features to vp.

The *vp* Hamiltonian can be defined similar to the band (co-operative) JT effect [3]. Likewise, we introduce electron and phonon ladder operators, as well as coupling constants and local energy terms. This defines our problem within the electron-phonon framework. (Details can be found elsewhere [4].) In so doing we see that similar logistics can be applied to many coupled fermion-boson systems, not only electron-phonon ones. We now extend the *vp* definition accordingly, as follows:

We consider an extended global Hamiltonian introducing ladder operators in 2$^{nd}$ quantization, $a_{n\alpha s}$ and $a_{n\beta s}^\dagger$, etc. for fermions, $b_{ns}$ and $b_{ns}^\dagger$ for bosons. Their mixing Hamiltonian reads

$$H_{mix} = \sum_{n\alpha\beta s} g_{n\alpha\beta s} (b_{ns}^\dagger + b_{ns}) (a_{n\alpha s}^\dagger a_{n\beta s} + a_{n\beta s}^\dagger a_{n\alpha s}) \quad (1)$$

where $g_{n\alpha\beta s} = G_{n\alpha\beta}(M_{n\alpha\beta s}\omega_{n\alpha\beta s}^2/h\nu_{n\alpha\beta s})^{-1/2}$ is the mixing constant, $M_{n\alpha\beta s}$ the reduced mass and $\nu_{n\alpha\beta s}$ the vibrational frequency of the coupled oscillator, α and β are two fermion orbital bands, nearly-degenerate and of opposite parities, s is the spin variable, n is the site index.

$a_{n\alpha s}^\dagger$ and $b_{ns}^\dagger$, etc. are superpartners. To obtain the complete global Hamiltonian, we add up the diagonal energy terms, fermion and boson, respectively:

$$H_f + H_b = \sum_{n\alpha\beta s} E_{n\alpha s}\, a_{n\alpha s}^{\dagger} a_{n\alpha s} + \sum_{n\alpha\beta s} (n + ½) h\nu_{n\alpha s}\, b_{ns}^{\dagger} b_{ns} \qquad (2)$$

as well as the fermion hopping (band) term

$$H_{fhop} = \sum_{n\alpha s} t_{n\alpha s}\, a_{n\alpha s}^{\dagger} (a_{n-1\alpha s} + a_{n+1\alpha s}) \qquad (3)$$

The complete global Hamiltonian then reads

$$H = H_f + H_b + H_{fhop} + H_{mix} \qquad (4)$$

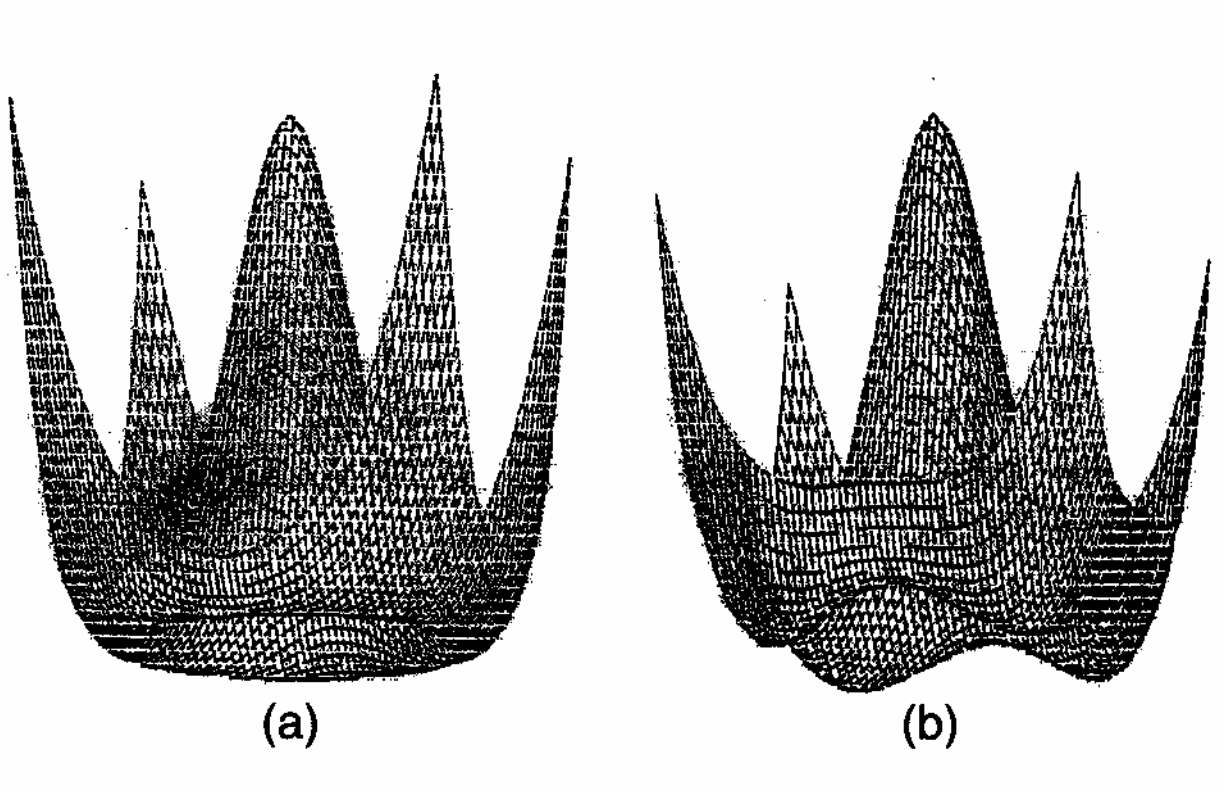

Figure 1: Local adiabatic energies of planar vibronic polarons to first order (a) and third order (b) of the electron-vibrational mode coupling constant. While the off-centered circle is clearly identifiable in (a), rotational barriers along the off-center ring are also seen adding up to (b).

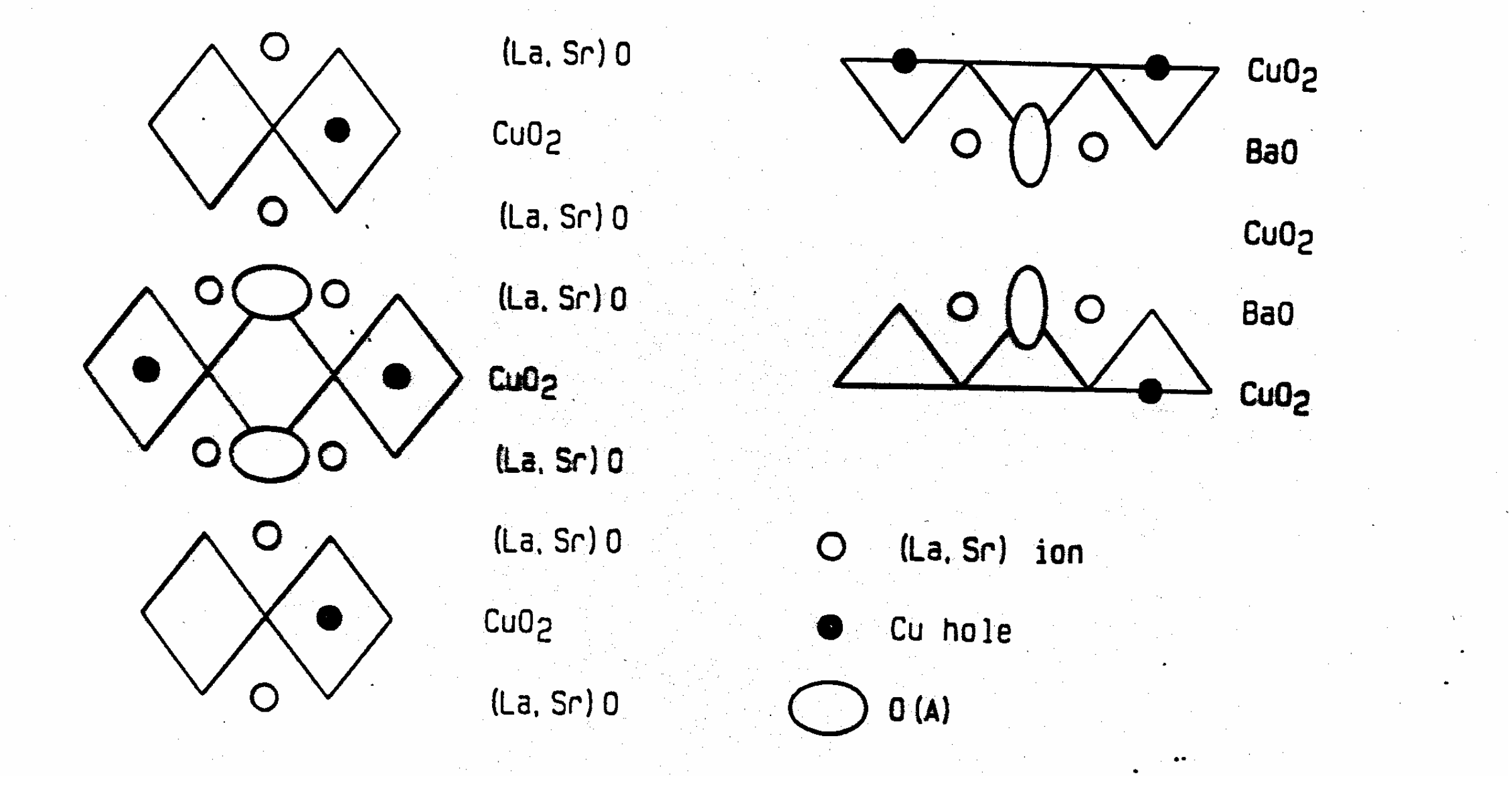

Figure 2: Conceived oblate (left) and prolate (right) off-center oxygen volumes due to appropriate mode coupling in high-Tc superconductors. Vertical is the c-axis of the crystal.

Because of the large asymmetry of masses in favor of the phonon, eigenvalues of (1) can be found approximately by means of the adiabatic approximation [5] which gives potential energies for *vp* rotation. One such is shown in Figure 1 for two coupling situations.

## 2. Off-center volume and deformed nuclei

Due to tunneling across the off-center to on-center barrier, the atomic particle which represents the distortion around a small *vp* polaron appears smeared within the off-center volume [6]. Depending on the symmetry of the coupled boson the latter volume may be oblate or prolate. In nuclear matter such may be the physics behind the deformed nuclei [7]. The asymmetric form of the deformed nuclei (boson coupling) or the smeared off-center ions (phonon coupling) is the most direct appearance of the shape of the off-centered species. For this reason, its experimental observation is a primary object of the *vp* study. Examples of conceived oblate and prolate forms of off-center volumes are shown in Figure 2.

## 3. Electrostatic *vp* polarizability

The vp has an associated off-center displacement of the center of charge which creates an electric dipole inherent for the polaron. It has also associated dipoles arising from the mixing of electronic states by the phonon coupling in accordance with the coupling Hamiltonian (1). The two kinds of dipoles are clearly interrelated. However, due to flip-flops across the centro-symmetric site the average electric dipole is vanishing because each flip-flop reverses to the opposite the sign of the dipole without altering its magnitude. As a result, *vp* is polarizable electrostatically like a polarizable atom. An immediate consequences of vp polarizability is, likewise, the occurrence of dispersive pair interactions between the species.

Now if $\alpha$ is the vp polarizability then the Van der Waals (VdW) pair interaction however complex it has been simplified to the form [8] $\frac{1}{2}\, h\nu_{rot}(\alpha\ /\kappa R^3)^2$ where $\nu_{rot}$ is the flip-flop frequency, R is the pair separation, $\kappa$ is an appropriate dielectric constant. The polarizability is related to the electrostatic dipole by way of $\alpha = p^2/\Delta t$, a second- order perturbation result [9]. For a typical PJT system the intralevel tunneling splitting $\Delta t = E_{gap} \exp(-2E_{JT}/h\nu)$, $\nu$ is the coupled bare vibrational frequency, $E_{gap}$ is the PJT interlevel gap energy. The current logistics also estimate the flip-flop frequency as $h\nu_{rot} \sim \Delta t$ [10].

The $\exp(-2E_{JT}/h\nu)$ factor (exponentially small) where $E_{JT}$ is Jahn-Teller's energy is known as Holstein's reduction factor [11]. It's entering in the denominator largely increases the polarizability $\alpha$ beyond the 2$^{nd}$ order perturbation estimate. At present we have no better way to accout for the polarizability though extensions are under way. It should be stressed that only even-order perturbation estimates are non-vanishing because the energy functional is at the extremum otherwise. A VdW molecule formed as explained above is shown in Figure 3 for a conceivable case in the LASCO superconductor.

## 4. Universal VdW binding

There have been suggestions that an universal molecular binding mechanism should be operative at the initial and later stages of the Universe to promote the formation of matter from individual atoms [12]. Here we stress that the universal binding proposal does not rule out the formation of light atoms from hydrogen atoms by hydrogen synthesis. Instead, they are rather assumed to complement each other. While synthesis is more or less clear from the viewpoint of cosmology, binding is not, as far less is known of the nature of the universal mechanism. Many have suspected the Van-der-Waals interaction but due to its being too weak, even though universal, there has been a widely spread skepticism about its role in the universe.

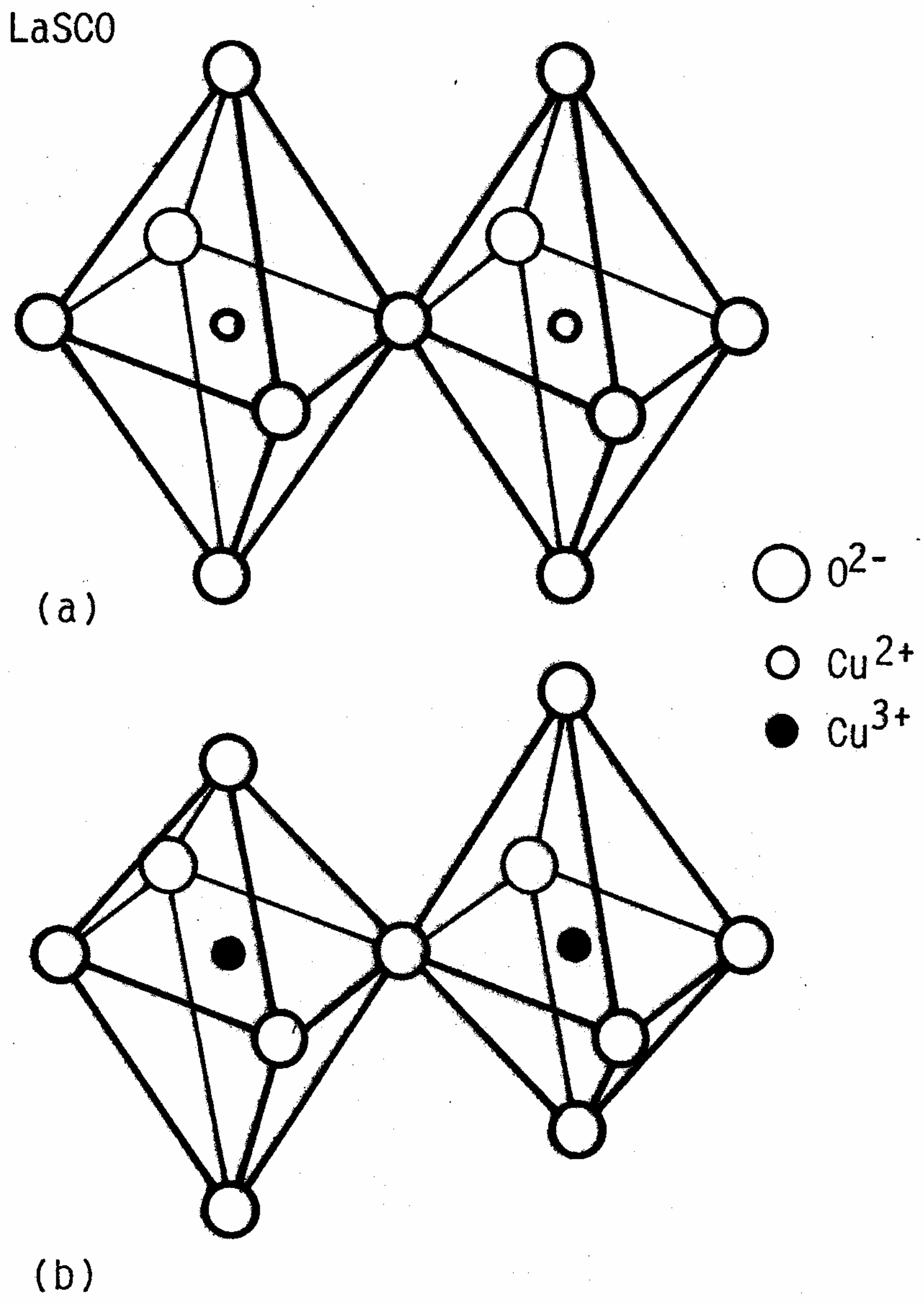

Figure 3: Conceivable *vp* pairs (bipolarons) for the high-$T_c$ $La_{2-x}Sr_xCuO_4$ superconductor.

But, what if the vp VdW interaction is controllable by varying some of its parameters making it stronger at the initial stages and weaker as the amount of primary atoms accumulates?

Indeed, as the amount of atoms formed by synthesis and of molecules by binding increases, so does the pressure in the region of a given molecular cluster. A WdW parameter that is particularly sensitive to pressure is the interlevel energy gap $E_{sp} = |E_p – E_s|$, the level spacing between an odd parity energy level $E_p$ and an even parity energy level $E_s$, familiar in the PJT theories. The expected pressure coefficient being positive for a bandgap [13], it leads to the conclusion that the VdW force while strong at the early stages will diminish in strength as the material pressure accumulates [14].

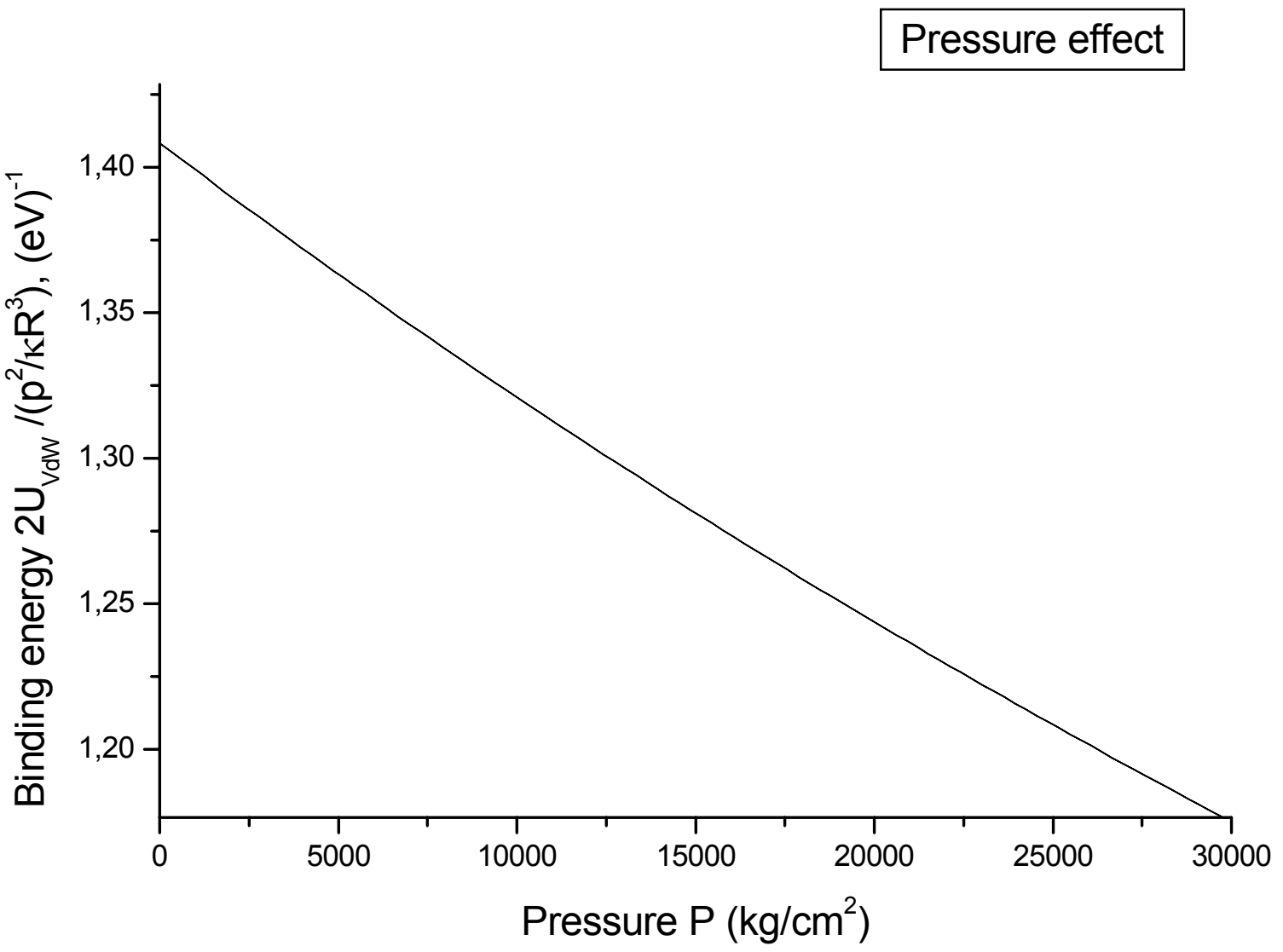

Figure 4: Calculated pressure effect on the colossal Van der Waals binding energy using data of Ref.13. The molecular system moves to the right as the universe evolves in time.

We do not know much of the nature of the original VdW species operative at the early stages. They may be sort of precursor molecules though they may have preserved the basic features of full-grown molecules. Indeed, as far as electrostatic polarization is concerned separate molecules may act as do species within a molecular crystal [15]. This seems to lay ground to our basic assumption concerning isolated and crystal-borne molecules in a molecular crystal. The missing universal binding is nothing but the strong polarization mechanism operating amongst the dilute low-pressure atmosphere of precursor molecules. The typical pressure dependence of the energy gap in semiconductors is shown in Figure 4 [13].

## 5. Breaking down of supersymmetry

In particle physics, supersymmetry (SUSY) is one that relates elementary particles of one spin to another particle that differs by half a unit of spin, the two particles being known as superpartners. In a supersymmetric theory, for every type of boson there exists a respective type of fermion, and *vice versa* . With electrons (fermions) and phonons (bosons) in mind as superpartners, the extended band vibronic Hamiltonian describes creation and annihilation processes which suggest that the ultimate configurational lowering by vibronic effects is related to the breakdown of supersymmentry. Considering the mixing term, e.g. [16]

$$(b_{n1}^{\dagger} + b_{n1})\,(a_{n\alpha+\frac{1}{2}}^{\dagger}\, a_{n\beta+\frac{1}{2}} + a_{n\beta+\frac{1}{2}}^{\dagger}\, a_{n\alpha+\frac{1}{2}}) =$$

$$(b_{n1}^{\dagger}\, a_{n\alpha+\frac{1}{2}}^{\dagger})\, a_{n\beta+\frac{1}{2}} + (b_{n1}^{\dagger} a_{n\beta+\frac{1}{2}}^{\dagger})\, a_{n\alpha+\frac{1}{2}} + a_{n\alpha+\frac{1}{2}}^{\dagger}\,(b_{n1} a_{n\beta+\frac{1}{2}}) + a_{n\beta+\frac{1}{2}}^{\dagger}\,(b_{n1} a_{n\alpha+\frac{1}{2}}),$$

where on the second row created or annihilated pairs are inserted between round brackets, the vibronic Hamiltonian describes the creation and annihilation of supersymmetric pairs at the expense of the respective annihilation and creation of single particles which implies the breakdown of supersymmetry since pairs are annihilated.

## 6. Chirality in *vp* propagation

Considering the conceivable quasilocal *vp* rotation in equatorial plane before moving to a neighboring plane, we come to the conclusion that the species moves along a helical path. Such free *vp* propagation is characterized by a finite orbital angular momentum related to the distance between neighboring circular segments. The wave function of a free *vp* therefore is $\psi(r) = H(\varphi)\exp(-ik_z z)$ where $H(\varphi)$ is Mathieu's planar function [6], $\varphi$ is the azimuth angle, z is the axial coordinate. The helical motion and attached chirality may be considered inherent to almost free carriers [17]. It is possible that vp has a definite chirality not any of the two, which may lead to parity nonconservation as it does for the neutrino [18]. However this can only be verified through experiment.

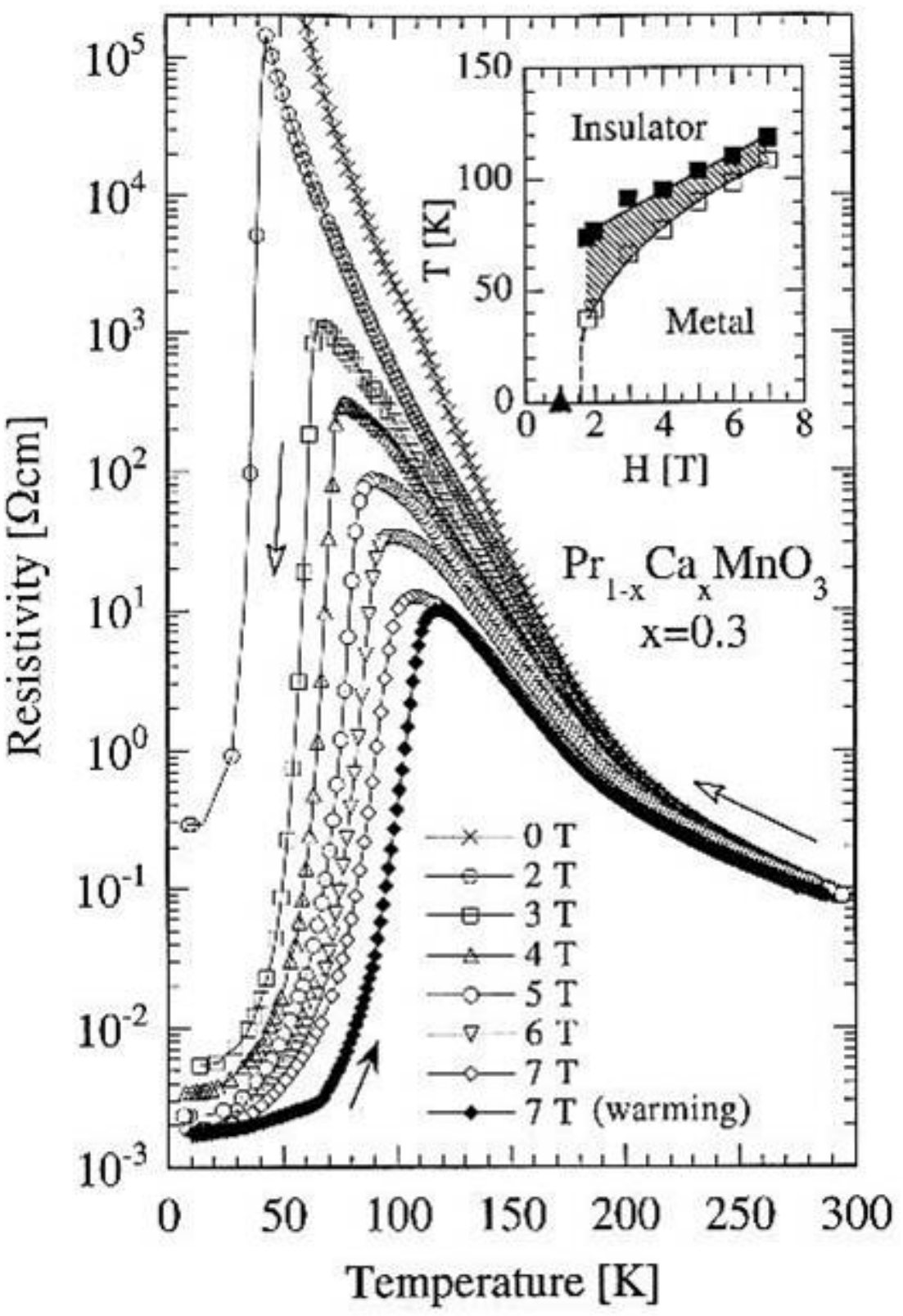

Figure 5: Experimental temperature dependences of electrical resistivity curves across the critical temperature showing both the low temperature low resistivity region (to the left of maximum) and the high temperature high resistivity region (to the right of maximum). Experimental data by Professor Tokura, Tokyo University. Courtesy Internet.

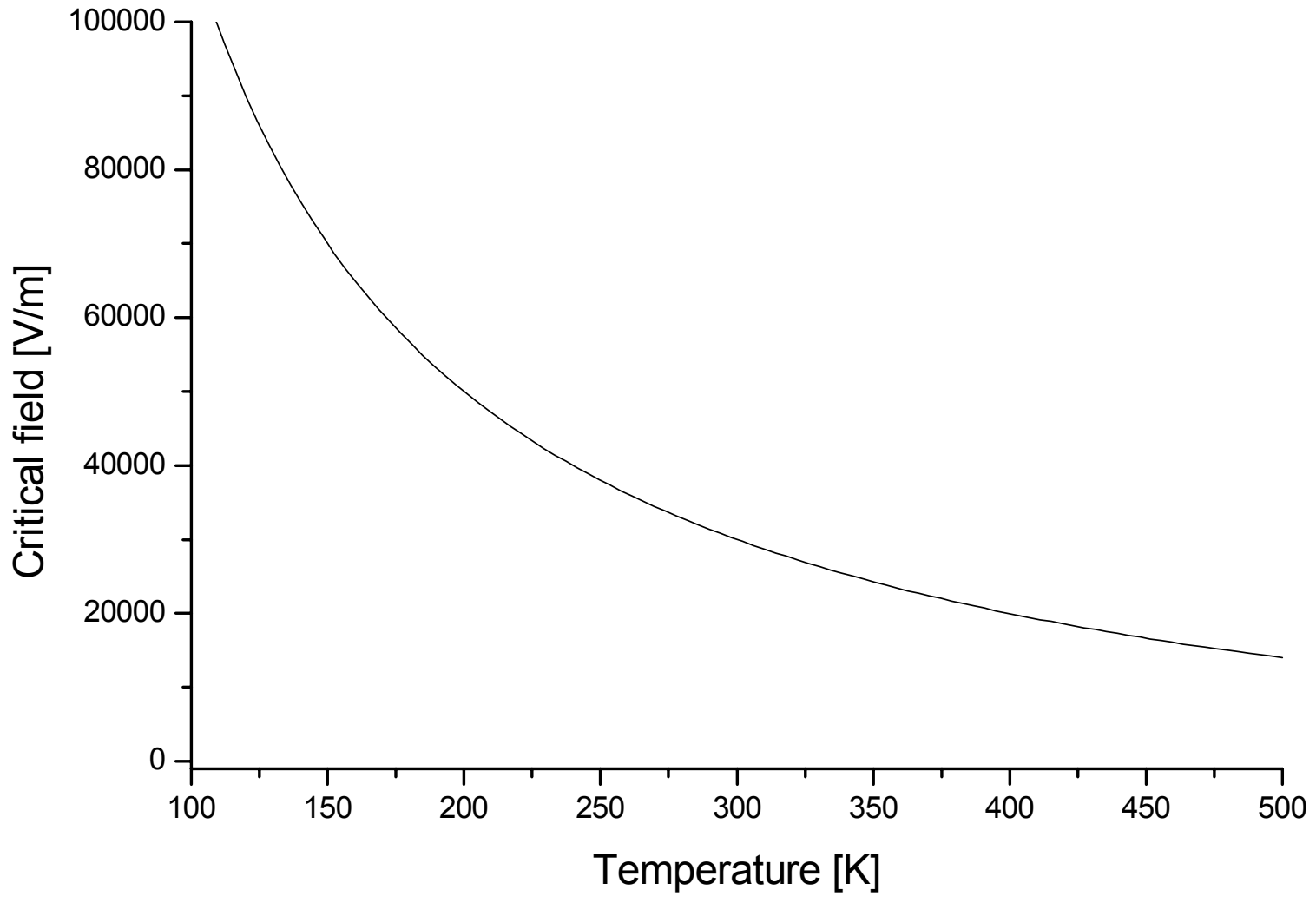

Figure 6: Theoretical temperature dependence of the critical field calculated by means of *vp* theory. This curve is to be compared with the graphic material shown in Figure 5(a). A close relation- ship is evidenced. From Ref.19.

## 7. Colossal magnetoresistance through orbital magnetic dipoles

The *vp* helical motion along the z-axis is accompanied by surging magnetic dipoles which may give rise to magnetic response of a *vp* medium. An appearance of that kind is the colossal magnetoresistance which occurs when the *vp* material is heated to make a crossover across some characteristic temperature $T_C$. We built up our description on the premise that in accordance with the present knowledge, the conduction of the medium is through large polarons below $T_C$. Above $T_C$ the polarons condense into a lattice of small vp characterized by high electric resistance. There is a borderline about $T_C$ which separates the two regions. The borderline has been calculated and found feasible [19], as shown in Figures 5 and 6.

## 8. Extension to other coupled fermion–boson system

We have thus far considered some features of the electron-phonon *vp* systems that may prove extendable to a wider appearance. Indeed, there seem to be a number of conceivable fermion-boson coupled systems that may be considered within the above context. Of these, the nucleon-π meson system appears plausible. However, due to the smaller boson mass (at least five times smaller than the nucleon's) no adiabatic approximation will be an useful premise. Likewise, an antiadiabatic approximation may also be of little practical merit too [16], so that numerical methods may come to the rescue.

Accounting for the above six vp appearances in electron-phonon systems, we consider them extended to more general fermion-boson systems and advise LHC experiment planners for the most immediate future. It seems that the relationship between (super)symmetry breaking and *vp* formation through coupled configurational symmetry lowering may be most appealing in experiments during later-initial stages following the big bang. During that time a strong universal binding mechanism may be highly effective to lay down nuclei for material cluster formation and related more intense accumulation of ordinary matter. Dark matter is also high on the agenda during those stages and it's formation may be related to both the universal mechanism and the symmetry breaking, as expected by many.

The form of *vp* propagation characterized by chirality may also be found highly instructive, particularly if it leads to the breaking of parity or to another symmetry too [17]. One is tempting to speculate that in such a case the left-right response of free space may be at stake, unless something like the parity nonconservation due to neutrino can be pointed out with confidence. Colossal magneto-resistance may also lead to excitement among the experimentalists, due to orbital dipoles at least as part of the story, and result in devising sensitive measuring units for external magnetic fields. We forsee bright prospects for common influence through interdisciplinary studies of fermion-phonon systems of whatever nature.

We have presently surveyed some basic *vp* features leaving aside others, such as self-trapping, optical spectra, etc. which have been dealt with in earlier publications [20].